\documentclass[cameraready]{Interspeech}
\usepackage{multirow}

\title{Samsone: A Family of Open Small Audio Language Models for On-Device Inference}

\author[affiliation={1,2},equalcontribution]{Piotr}{Masztalski} 
\author[affiliation={1},equalcontribution]{Michal K.}{Grzeszczyk}
\author[affiliation={1}]{Olaf}{Sikorski}

\address{
    $^1$ Samsung R\&D Institute Poland\\
    $^2$ AGH University of Kraków, Poland
}

\email{\{p.masztalski, m.grzeszczyk, o.sikorski\}@samsung.com}


\usepackage{comment}
\usepackage{hyperref}

\begin{document}

\maketitle

\begin{abstract}
    The success of Large Audio Language Models has driven the development of massive multimodal networks exceeding billions of parameters. However, the demand for privacy-preserving, low-latency processing has shifted focus toward Small Audio Language Models (SALMs) capable of on-device execution. In this paper, we introduce Samsone, a family of SALMs designed for edge computing. Our core model, Samsone-134M, establishes a new state-of-the-art for its size class across multiple benchmarks. We further explore the scaling laws of SALMs by introducing Samsone-99M and Samsone-356M. Despite their compact footprint, the Samsone family delivers performance competitive with models orders of magnitude larger. To foster open research and reproducibility, we train Samsone on publicly available data. We release the training code, model weights, mobile-optimized checkpoints and provide an open-source Android application to demonstrate real-time on-device inference of Samsone.
\end{abstract}

\section{Introduction}
Traditionally, audio tasks such as sound classification \cite{gemmeke2017audio}, audio captioning \cite{kim2019audiocaps,drossos2020clotho}, and speaker identification \cite{nagrani2017voxceleb} were addressed by task-specific architectures \cite{elizalde2023clap}. Recently, the rapid advancement of Large Language Models (LLMs) has catalyzed the development of unified Large Audio Language Models (LALMs) \cite{kong2024audio,ghosh2025af2,ghosh2025af3,comanici2025gemini,hurst2024gpt,li2025reinforcement, gong_ltuas,gong2024listen}. Using audio-text pairs to train in the Audio Question Answering (AQA) setting, these models achieve generalized performance across diverse tasks. While the audio domain initially faced data scarcity, the use of multimodal LMs to generate synthetic, high-reasoning datasets, such as OpenAQA \cite{gong2024listen}, ReasonAQA \cite{deshmukh2025mellow}, and AudioSkillsXL \cite{ghosh2025af3}, has significantly boosted LALMs capabilities.

While the performance gains of scaling LLMs are undeniable, they come with significant environmental and computational costs \cite{singh2025survey}. Furthermore, strict privacy requirements in domains such as healthcare \cite{ong2024ethical}, the need for offline processing, and the rise of agentic frameworks for specialized, repetitive tasks \cite{belcak2025small} have created a demand for small-scale models. 

\begin{figure}[t]
    \centering
    \includegraphics[width=8cm]{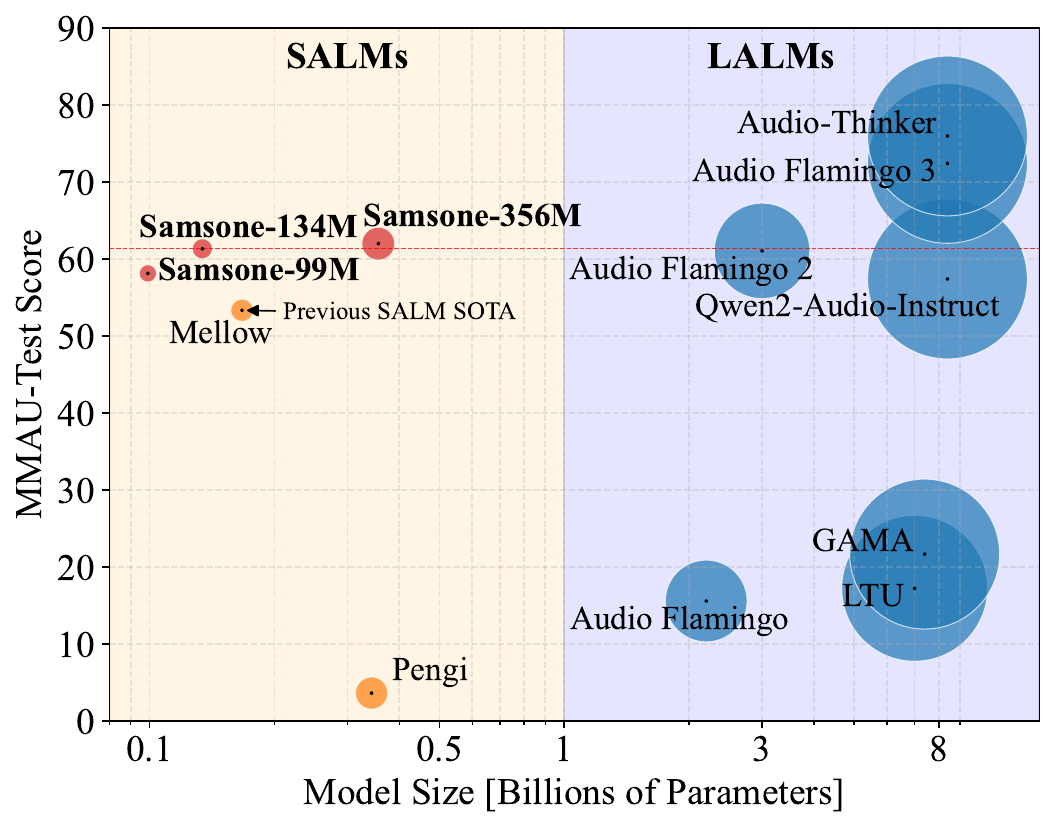}
    \caption{Samsone establishes the new state of the art on the MMAU benchmark among SALMs. Both Samsone-99M and Samsone-134M beat the previous SOTA (Mellow) while containing less parameters. All of the Samsone variants are also competitive with much larger, multi-billion parameter LALMs.}
    \label{fig:model_performance}
\end{figure}

Defining the boundary between small and large models remains ambiguous. In the text domain, where parameter numbers frequently exceed hundreds of billions (e.g., Qwen3-235B \cite{yang2025qwen3}), even multi-billion parameter architectures like Llama3-8B \cite{grattafiori2024llama} are categorized as small. However, the audio domain operates on a different scale: state-of-the-art (SOTA) LALMs rarely exceed 10 billion parameters, with leading models such as Audio Flamingo 3 (8.4B) \cite{ghosh2025af3}, typically ranging between 3B and 9B. Consequently, in this paper, we define Small Audio Language Models (SALMs) as those with fewer than 1 billion parameters.

Existing SALMs remain underrepresented. Pengi (323M) \cite{deshmukh2023pengi} pioneered treating audio tasks as text generation but struggled with complex reasoning due to limited training text diversity. This was partially addressed by Mellow (167M) \cite{deshmukh2025mellow}, which utilized the ReasonAQA dataset. Despite their potential for on-device execution, actual mobile deployment and benchmarking of such models remain largely unexplored.

\begin{figure*}[t]
    \centering
    \includegraphics[width=16.5cm]{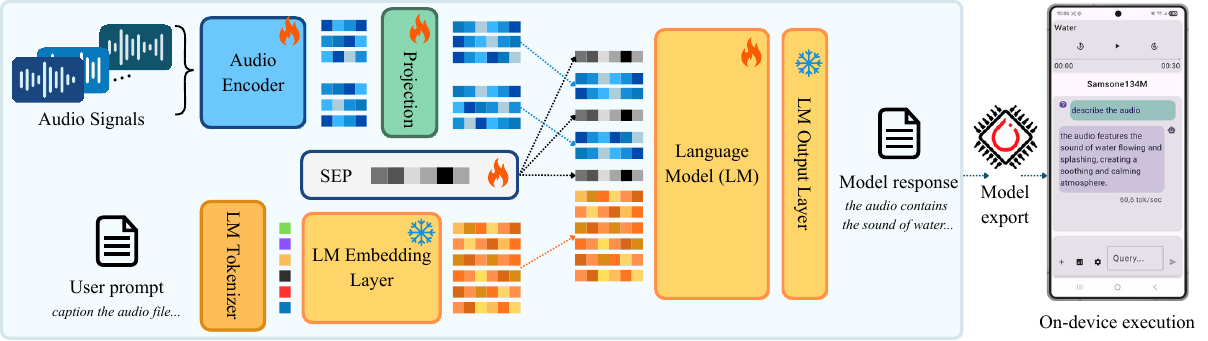}
    \caption{Samsone family of SALMs consisting of Audio Encoder, projector and Language Model with on-device execution example.}
    \label{fig:system_architecture}
\end{figure*}

In this paper, we introduce Samsone, a family of SALMs that achieve SOTA performance across various audio tasks (Fig. \ref{fig:model_performance}). Samsone adopts a standard ALM architecture, comprising an Audio Encoder (AE), a modality projector mapping audio embeddings to the text embedding space, and an LM text decoder. Through rigorous dataset curation and token pruning, we developed Samsone-134M a 134-million-parameter model that outperforms existing SALMs such as Pengi and Mellow on diverse audio benchmarks. We further explore the efficiency frontier by depth-pruning the LM to create Samsone-99M, a sub-100M parameter model that maintains competitive performance despite its compact size. Conversely, we scale up the architecture to create Samsone-356M, which leverages a larger LM backbone to deliver enhanced accuracy. Finally, we validate the practical utility of the Samsone family by exporting model weights and demonstrating real-time inference on a commodity smartphone.

The main contributions of this work are: 1. We introduce Samsone-134M, the best performing ALM in its class. It achieves SOTA results within its size range and remains competitive with models orders of magnitude larger. 2. We present Samsone-99M and Samsone-356M to explore the scalability of SALMs across varying on-device computational constraints, providing insights into the trade-offs between model capacity and efficiency. 3. To promote reproducibility and accelerate research in the field, we open-source the Samsone Family. This includes the training code, server-side and on-device weights, as well as an Android application for inference on smartphones\footnote{https://github.com/SamsungLabs/samsone}.

\section{Proposed Method}
In this section we introduce Samsone - its architecture, multiple size variants and design choices for efficient operation.

\subsection{Model architecture}

Samsone is a multimodal language model (LM) that accepts audio and text inputs to produce text output (Fig. \ref{fig:system_architecture}). It adopts a standard ALM architecture with an audio encoder, a modality projector that maps audio embeddings into the text embedding space, and a text backbone that processes the multimodal input.

As our audio backbone, we extract the encoder part of Whisper \cite{radford2023robust}, a transformer-based model originally designed for Automatic Speech Recognition. The audio input gets processed through the Whisper encoder and is subsequently passed to the non-linear modality projector. Its role is to align the feature dimension of raw AE outputs with the text embedding dimension expected by the LM. The sequence length of the projection module output is equal to the sequence length of its input, the audio embeddings.

For the language modeling task, we choose the SmolLM2 \cite{allal2025smollm2} models in 135M and 360M sizes, depending on the Samsone variant. SmolLM2 is a decoder-only LM based on the LLaMA2 \cite{llama} architecture trained on a highly curated multi trillion token text dataset. The text inputs are initially passed through the SmolLM tokenizer, resulting in a sequence of tokens that is mapped into a corresponding sequence of text embeddings by the SmolLM2 embedding layer.

The audio-text input to the SmolLM2 transformer blocks is constructed, by concatenating the projected audio embeddings with the text embeddings. As Samsone is designed to handle an arbitrary number of audio signals, we additionally introduce a trainable SEP token embedding that separates the input audio embeddings from each other, and from the surrounding text embeddings. This enables Samsone to handle multiple variable-length audio inputs while preserving the information about the number of audio signals in each input sequence and to position the audio embedding sequences in any part of the audio-text input. The resulting multimodal input is then processed by the remaining part of the SmolLM2 model to produce a text output.

\subsection{Size optimizations}

We reduce the model size by targeting two components: (i) the embedding matrix, and (ii) the number of Transformer layers.

\noindent \textbf{Vocabulary reduction (VR):} In LMs, the input text is tokenized and each token is mapped to a continuous embedding vector. When the embedding dimension is relatively large (e.g., 576 in SmolLM2-135M) and the vocabulary is extensive (49152 tokens), the embedding matrix constitutes a significant fraction of the total parameters. In SmolLM2-135M, the embedding layer alone contains more than 28M parameters (approximately 21\% of the total model size). To reduce this overhead, we constrain the training text to lowercase ASCII characters (as in \cite{deshmukh2025mellow}) and filter out rarely useful tokens. Specifically, we remove tokens with more than four whitespace characters or more than three special characters. This step enables us to eliminate 15042 tokens from the vocabulary, reducing the embedding matrix by 8.7M parameters while preserving coverage of the target domain.

\noindent \textbf{Depth pruning (DP):} Our second optimization strategy focuses on reducing model depth \cite{kim2024shortened,sandri2025ssp}. Modern LMs are composed of multiple stacked Transformer blocks, each contributing substantially to the total parameter count. In our setup, all Transformer blocks are trainable, which allows us to apply straightforward depth pruning. Removing a single Transformer block in SmolLM2-135M reduces the model size by approximately 3.5M parameters. This simple yet effective strategy provides a controllable trade-off between model capacity and footprint, enabling systematic scaling.

\subsection{Model variants}
\begin{table}[t]
    \caption{Parameter count comparison of Samsone size variants with vocabulary reduction (VR) and depth pruning (DP).}
    \label{tab:model_variants}
    \centering
    \eightpt
    \begin{tabular}{@{}lcccccc@{}}
        \toprule
        \textbf{Variant} & VR & DP & \textbf{Enc.} & \textbf{Proj.} & \textbf{LM} & \textbf{Total size} \\ 
        \midrule
        Samsone-99M  & \checkmark & \checkmark & 9M & 0.5M & 90M & 99M \\
        Samsone-134M & \checkmark & $\times$   & 9M & 0.5M & 125M & 134M \\
        Samsone-356M & \checkmark & $\times$   & 9M & 1M & 347M & 356M\\
        \bottomrule
    \end{tabular}
\end{table}

Leveraging the described size optimizations, we introduce the Samsone family, consisting of three distinct variants. All models share a common AE based on the Whisper-Tiny architecture \cite{radford2023robust}. The core of the family is Samsone-134M, which utilizes the SmolLM2-135M backbone \cite{allal2025smollm2}. By applying our VR strategy, we decrease the total parameter count to approximately 134 million. To explore the lower bounds of on-device efficiency, we introduce Samsone-99M. This variant combines VR with DP, truncating the LM from 30 to 20 Transformer blocks by removing the last 10 layers to achieve a sub-100M parameter footprint. Finally, we scale the architecture upward with Samsone-356M, which employs the SmolLM2-360M backbone to provide higher reasoning capacity for more complex audio queries. The detailed parameter distribution for each variant is summarized in Table \ref{tab:model_variants}.

\section{Implementation details}
In this section, we describe the architecture of the Samsone family models, initialization strategy, datasets, training configuration, and the export procedure for on-device deployment.

\subsection{Models}
We initialize the AE from \textit{openai/whisper-tiny}, and the LM from \textit{HuggingFaceTB/SmolLM2-135M} (and \textit{SmolLM2-360M} for Samsone-356M), using the Transformers library \cite{wolf-etal-2020-transformers}. To bridge the audio and language representations, we adopt the nonlinear projector architecture introduced in Mellow \cite{deshmukh2025mellow}. The projector consists of two linear layers with a GeLU activation in between, followed by a residual connection and a final layer normalization step. The AE produces frame-level audio embeddings, which we temporally average-pool to obtain a fixed-length representation of 50 audio tokens per sample. To explicitly separate audio and text modalities, we insert a trainable SEP token before, after, and between audio tokens. This design helps the LM distinguish modalities and improves their alignment.

\subsection{Data}
We train all Samsone variants on two datasets: ReasonAQA \cite{deshmukh2025mellow} and AudioSkillsXL \cite{ghosh2025af3}. AudioSkillsXL is a large-scale dataset comprising 8 million question–answer pairs across sound, music, and speech. ReasonAQA is designed specifically for reasoning and contains 1 million QA pairs, many of which require reasoning over two audio inputs. We identified a strong class imbalance in the multiple-choice subset of ReasonAQA: option (b) is the correct answer in the majority of examples, exceeding the combined frequency of all other answers. This imbalance can bias the model toward favoring the second option. To mitigate this issue, we randomly permute answer choices during training to enforce a uniform distribution of correct answers. As SmolLM base models do not utilize prompt templates, we append a postfix string \textit{" answer: "} to the prompt (before the answer) to explicitly delimit the model response.

\subsection{Training setup}
We implement all models in PyTorch and use pytorch-lightning \cite{Falcon_PyTorch_Lightning_2019} to manage the training pipeline. Given the relatively small parameter count of Samsone models, all components except the LM embedding layer are trainable. We experimented with multi-stage training strategies commonly used in LALMs \cite{ghosh2025af2,ghosh2025af3}, but observed no performance gains. Therefore, we adopt a single-stage training procedure. Each model is trained for 100 epochs on a single NVIDIA RTX PRO 6000 Blackwell 96GB GPU, with each epoch consisting of 200,000 training examples. We use the AdamW optimizer \cite{loshchilov2017decoupled} with a cosine annealing learning rate scheduler. The schedule includes 10 linear warm-up epochs and a minimum learning rate of 1e-7. The learning rate is set to 3e-4 for Samsone-99M and Samsone-134M, and 1e-4 for Samsone-356M. Models are trained using standard token-level cross-entropy loss. After training, we export checkpoints with XNNPACK using ExecuTorch for on-device inference.

\section{Experiments and results}
In this section we describe our experiments on various audio tasks including audio understanding, captioning and AQA. We present the ablation study of key components of Samsone-134M and benchmark our models on a Samsung Galaxy S25 Ultra. To ensure reproducible results, in all of the following experiments, we opt for the greedy text decoding strategy during inference.

\subsection{Massive Multitask Audio Understanding (MMAU)}

\setlength{\tabcolsep}{4pt}

\begin{table*}[th]
  \caption{Comparison of Audio Language Models performance on MMAU and MMAU-Pro datasets. \cite{sakshi2024mmaumassivemultitaskaudio}.} 
  \label{tab:results_mmau}
  \centering
  \eightpt
  \begin{tabular}{l c cc cc cc cc c}
    \toprule
    & & \multicolumn{2}{c}{\textbf{Sound}} & \multicolumn{2}{c}{\textbf{Music}} & \multicolumn{2}{c}{\textbf{Speech}} & \multicolumn{2}{c}{\textbf{Avg.}} & \multirow{2}{*}{\textbf{MMAU-Pro}} \\
    \cmidrule(lr){3-4} \cmidrule(lr){5-6} \cmidrule(lr){7-8} \cmidrule(lr){9-10}
    \textbf{Name} & \textbf{Size} & Test-mini & Test & Test-mini & Test & Test-mini & Test & Test-mini & Test \\
    \midrule
    \multicolumn{11}{c}{\textbf{Large Audio Language Models}} \\
    \midrule
    LTU \cite{gong2024listen}	              & 7B   & 20.42 & 20.67 & 15.97 & 15.68 & 15.92 & 15.33 & 17.44 & 17.23 & 33.46\\
    GAMA \cite{ghosh2024gama}	              & 7.4B & 31.83 & 30.73 & 17.71 & 17.33 & 12.91 & 16.97 & 20.82 & 21.68 & 33.20\\
    SALMONN	\cite{tang2024salmonn}	          & 13B  & 41.14 & 42.10 & 37.13 & 37.83 & 26.43 & 28.77 & 34.90 & 36.23 & 39.60\\
    Qwen2-Audio-Instruct \cite{chu2024qwen2}  & 8.4B & 67.27 & 61.17 & 56.29 & 55.67 & 55.26 & 55.37 & 59.60 & 57.40 & 45.41\\
    GPT-4o Audio \cite{hurst2024gpt}          & --   & 64.56 & 63.20 & 56.29 & 49.93 & 66.67 & 69.33 & 62.50 & 60.82 & \textbf{52.50}\\
    Audio Flamingo 2 \cite{ghosh2025af2}      & 3B   & 71.47 & 68.13 & 70.96 & 70.20 & 44.74 & 44.87 & 62.40 & 61.06 & 42.60\\
    Audio Flamingo 3 \cite{ghosh2025af3}      & 8.4B & \textbf{79.58} & \textbf{75.83} & \textbf{73.95} & \textbf{74.47} & \textbf{66.37} & \textbf{66.97} & \textbf{73.30} & \textbf{72.42} & 51.70\\
    \midrule
    \multicolumn{11}{c}{\textbf{Small Audio Language Models}} \\
    \midrule
    Pengi \cite{deshmukh2023pengi} & 323M               &  3.00 &  3.87 &  0.29 &  2.27 &  3.30 &  4.77 &  2.20 &  3.63 & 28.61\\
    Mellow \cite{deshmukh2025mellow} & 167M             & 69.37 & 66.17 & 56.89 & 58.13 & 30.63 & 35.73 & 52.30 & 53.34 & 27.50\\
    \textbf{Samsone-99M (ours)} & 99M                   & 72.97 & 71.13 & 60.78 & 61.17 & 37.84 & 42.10 & 57.20 & 58.13 & 36.83\\
    \textbf{Samsone-134M (ours)} & 134M & \textbf{76.28} & \textbf{73.23} & \textbf{66.47} & \textbf{62.87} & \textbf{46.25} & \textbf{47.90} & \textbf{63.00} & \textbf{61.33} & \textbf{37.57}\\
    \midrule
    \textbf{Samsone-356M (ours)} & 356M & 75.98 & 74.27 & 70.34 & 65.83 & 44.74 & 45.90 & 63.70 & 62.00 & 40.67\\
    
    \bottomrule
  \end{tabular}
\end{table*}



We assess Samsone's audio understanding abilities using the MMAU benchmark, which includes 10,000 human-annotated AQA pairs across three domains: speech, sound, and music \cite{sakshi2024mmaumassivemultitaskaudio}. MMAU is designed to require expert-level knowledge and complex reasoning. Evaluation results presented in Table \ref{tab:results_mmau} show a strong improvement of all Samsone variants over the previous SALM SOTA, Mellow, across all domains. Even our smallest model, Samsone-99M, consistently outperforms Mellow, while containing over 40\% less parameters. It is worth mentioning, that we compare our solution to an improved version of Mellow (\textit{v0\_s})\footnote{https://github.com/soham97/mellow}, that was made available after the original paper publication.
Additionally, Samsone-134M remains competitive with much larger Audio Flamingo 2, beating the likes of 60x larger Qwen2-Audio or 100x larger SALMONN.

To evaluate our models in the most challenging environment, we utilize MMAU-Pro \cite{kumar2025mmau}, a benchmark with 5,305 AQA instances testing long-form audio comprehension, spatial audio reasoning and multi-audio understanding. We observe an even bigger improvement of Samsone-99M and Samsone-134M models (34\% - 36\%) over Mellow (Table \ref{tab:results_mmau}). We attribute this performance gain to our more diverse and extensive training data mix. Comparing Samsone to competing LALMs, we find that larger model size provides a slight advantage on this benchmark, which we argue, stems from greater inbuilt knowledge of LLMs, on which these LALMs are based on. Finally, there is a correlation of Samsone's performance gain with its size increase, which is consistent with scaling laws \cite{kaplan2020scaling}.

\subsection{Question Answering, Captioning and Reasoning}

To obtain a full picture of Samsone's performance, we test our models on audio captioning, simple AQA (yes/no and one word answers), and audio entailment tasks using Clotho and AudioCaps based datasets. In Table \ref{tab:results_audio_tasks}, we observe, that when it comes to simple AQA, all Samsone models outperform the competing SALMs. Similar holds true for the audio entailment tasks with a slight performance downgrade for Samsone-99M on AudioCaps. These results prove that Samsone has a particularly good grounding in audio while using very few parameters. Samsone models are also competitive on audio captioning tasks, beating Mellow on the Clotho dataset. We attribute the inferior SPICE score on AudioCaps to the lower share of the AudioCaps dataset in the full training dataset, than originally used for Mellow.  

\begin{table}[t]
    \centering
    \caption{Comparison of SALMs on audio captioning task on AudioCaps \cite{kim2019audiocaps} (AC) and Clotho \cite{drossos2020clotho} (CL)  with SPICE \cite{anderson2016spice} metric (sp.), on AQA task (ClothoAQA dataset \cite{lipping2022clotho} - (AQA)), and audio entailment task (Clotho\_CLE, AudioCaps\_ACE \cite{deshmukh2025audio}).}
    \label{tab:results_audio_tasks}
    \eightpt
    \begin{tabular}{lrrrrr}
        \toprule
        \textbf{Model} & \textbf{AC} & \textbf{CL} & \textbf{AQA} & \textbf{CLE} & \textbf{ACE}\\
                       & \textit{sp.} & \textit{sp.} & \textit{acc.} & \textit{acc.} & \textit{acc.} \\
        \midrule
        Pengi (323M) \cite{deshmukh2023pengi}     & 12.7 & 7.0  & 63.6 & 37.3 & 38.7\\
        Mellow (167M) \cite{deshmukh2025mellow}   & \textbf{17.8} & 9.4  & 71.4 & 91.2 & 89.7\\
        \textbf{Samsone-99M (ours)}               & 14.4 & 11.1 & 71.8 & 92.4 & 89.1\\
        \textbf{Samsone-134M (ours)}              & 14.4 & \textbf{11.6} & \textbf{73.8} & \textbf{93.4} & \textbf{93.7}\\
        \midrule
        \textbf{Samsone-356M (ours)} & 14.9 & 11.7 & 74.5 & 93.7 & 93.5 \\
        \bottomrule
    \end{tabular}
\end{table}

\subsection{Ablation Study}
We evaluate the key components of Samsone-134M architecture through an ablation study on MMAU benchmark. To ensure a controlled comparison, we utilize the LM without VP as the baseline, hence the 143M size (Samsone-134M-NP). We test the impact of replacing chosen modules with common alternatives: the AE with AST (Audio Spectrogram Transformer) \cite{gong21b_interspeech}, the projector with a linear layer, and the LM with GPT-2 \cite{radford2019language}. Results in Table \ref{tab:ablation} show that all alternative configurations yield inferior performance. These findings validate our architectural choices for balancing size and audio understanding capability.

\begin{table}[th]
    \caption{Ablation study of Samsone-134M.}
    \label{tab:ablation}
    \centering
    \eightpt
    \begin{tabular}{lcccc}
        \toprule
        \textbf{Model/change} & \textbf{Size} & \textbf{MMAU$_{mini}$} & \textbf{MMAU} \\ 
        \midrule
        \textbf{Samsone-134M-NP} & 143M & \textbf{61.80} & \textbf{59.92} \\
        \midrule
        w/ AST AE & 221M &  59.70 & 57.47 \\
        w/ GPT-2 LM & 133M & 58.80 & 57.96 \\
        w/ Linear projector & 142M  & 61.20 & 58.82 \\
        \bottomrule
    \end{tabular}
\end{table}

\subsection{On-Device Inference}
We develop an Android application to benchmark Samsone's inference latency on a Samsung Galaxy S25 Ultra CPU using 15 audio-query pairs. As shown in Table \ref{tab:on_device}, the generation process involves prefilling the LM cache with audio and query embeddings followed by autoregressive token generation. The Samsone family achieves generation speeds between 39 and 125 tokens per second depending on model size, confirming their suitability for real-time edge applications. Notably, these results were achieved without hardware-level optimizations, suggesting significant potential for further speed improvements.

\begin{table}[t]
    \caption{On-device inference of the Samsone family  models on Samsung Galaxy S25 Ultra.}
    \label{tab:on_device}
    \centering
    \eightpt
    \begin{tabular}{lccc}
        \toprule
        \textbf{Variant} & \textbf{Audio} & \textbf{Query} & \textbf{Generation}\\
                         & [ms]   & [ms]   & [tok/s]\\ 
        \midrule
        Samsone-99M  & 667 & 15 & 125 \\
        Samsone-134M &  757 & 23 & 87 \\
        Samsone-356M & 1116 & 47 & 39 \\
        \bottomrule
    \end{tabular}
\end{table}

\section{Conclusion}
In this paper, we introduced Samsone, a family of SALMs for on-device inference. Our core model, Samsone-134M, establishes a new SOTA for its size class, outperforming existing SALMs on multiple benchmarks, including MMAU (+15\%) and MMAU-Pro (+36\%), despite a 20\% reduction in parameter count. Samsone-134M also exceeds the performance of much larger models, such as GAMA and LTU, on these tasks. To facilitate further research we open-source all project artifacts. These include the training code, checkpoints, mobile-optimized weights, and an Android application for real-time inference. 

Our work has limitations. First, extensive AQA fine-tuning causes the LM to lose general-purpose linguistic capabilities. Second, we focused on parameter count as a proxy for efficiency, however, recent Precision-Aware Scaling Laws \cite{kumar2025scaling} suggest that larger, quantized models may offer superior performance-to-memory trade-offs. Finally, we have not yet implemented hardware-specific optimizations for mobile GPUs or NPUs. This will be explored in future research.

\section{Generative AI Use Disclosure}

The content of this paper was conceived, researched, and written entirely by the authors. Generative AI tools were utilized solely for editorial and grammatical refinement purposes, such as improving clarity, readability, and linguistic accuracy. These tools did not contribute to the generation of ideas, data analysis, interpretation of results, or creation of original content. The final manuscript reflects the full responsibility and intellectual effort of the authors.

\bibliographystyle{IEEEtran}
\bibliography{mybib}

\end{document}